\documentclass[pra,twocolumn,showpacs,preprintnumbers,amsmath,amssymb]{revtex4}
\usepackage{graphicx}
\usepackage{amssymb}
\usepackage{amsmath}

\usepackage{color,soul}

\begin{document}

\draft
\title{What role does the third law of thermodynamics play in Szilard engines?}
\author{Kang-Hwan Kim$^1$ and Sang Wook Kim$^{2}$}
\email{swkim0412@pusan.ac.kr}
\affiliation{$^1$Department of Physics, Pusan National University, Busan 609-735, Korea}
\affiliation{$^2$Department of Physics Education, Pusan National University, Busan 609-735, Korea}
\date{\today}

\begin{abstract}

The role of the third law of thermodynamics in the Szilard engine has been addressed. If the ground state is non-degenerate, the entropy production defined as the work extractable from the engine divided by temperature vanishes as temperature approaches zero due to the third law. The degenerate ground state induced by the symmetry or by accidence gives rise to non-zero entropy production at zero temperature associated with the residual entropy. Various physical situations such as the SZE consisting of bosons or fermions either with or without interaction have been investigated.

\end{abstract}


\maketitle
\narrowtext

\section{Introduction}
\begin{figure}
 \vspace*{1.8cm}
 \hspace*{-1.7cm}
 \includegraphics[width=12cm]{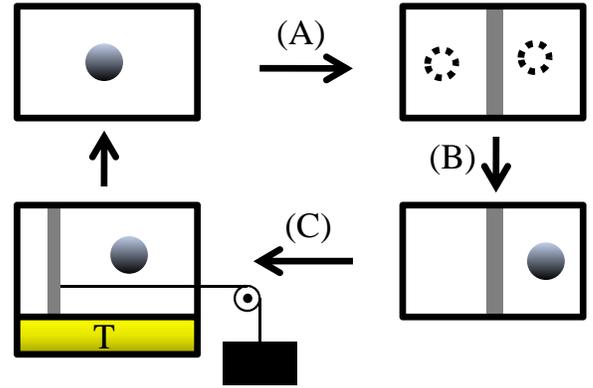}
 \vspace{-5.5cm}
  \caption{(Color online) Schematic diagram of the thermodynamic processes of the SZE. Initially a single atom is prepared in an isolated box. (A) A wall denoted as a vertical grey bar is inserted to split the box into to two parts. The atom is represented by the dotted circles to reflect that we are lack of the information on which side it is before the measurement. (B) The measurement is performed, and one acquires the knowledge of where the atom is. (C) A load denoted by a filled rectangle is attached to the wall to extract a work via an isothermal expansion at temperature $T$. \label{fig1}}
\end{figure}

Based upon the second law of thermodynamics Carnot proposed that every cyclic engines should consist of at least two reservoirs with two distinct temperatures, namely high and low temperature reservoirs. The former supplies entropy accompanying heat into the engine, while the latter absorbs the supplied entropy to eliminate it from the engine. It has been shown, however, that the engine proposed by Szilard \cite{Szilard29}, referred to as the Szilard engine (SZE), can be operated by using only a single reservoir, so that it was regarded as the so-called Maxwell's demon \cite{Leff03} which is believed to violate the second law of thermodynamics.
The thermodynamic cycle of the SZE consists of three steps as shown in Fig.~\ref{fig1}; (A) to insert a wall so as to divide a box into two parts, (B) to perform measurement to obtain information on which side the atom is in, and (C) to attach a weight to the wall to extract work via isothermal expansion with a thermal reservoir of temperature $T$ contacted.
Now it is widely accepted that the SZE does not contradict the second law since the {\em information} entropy of the SZE is transferred to the reservoir via measurement and erasure mediated by the demon \cite{Leff03,Brillouin51,Landauer61,Bennett82,Maruyama09}. The second law thus forces us to directly relate information to physical entropy, which reveals a profound role of information in nature. The SZE has been exploited to understand many physical problems in various contexts \cite{Scully03,Kim05,Raizen05} and realized in experiment \cite{Serreli07,Thorn08,Price08,Toyobe10}.

The third law of thermodynamics, proposed by Nernst in $1906$ \cite{Nernst06}, asserts that the entropy of a substance approaches zero as temperature approaches absolute zero. This statement is true only if the ground state of the substance is unique. Otherwise, the entropy approaches non-zero finite value called as the residual entropy. Many quantum statistical behavior in low temperature, e.g. the specific heat of solids, is governed by the third law. However, it is now found that the third law is a derived law rather than a fundamental law. Interestingly there is no study on the relation between the information and the third law. Moreover, it has never even been asked what the role of the third law is in the context of the information heat engine, especially the SZE. In the classical SZE, the entropy production defined as the work extracted from the engine divided by temperature is independent of temperature.

In this paper we show that the third law of thermodynamics manifests itself in the SZE, particularly in the low temperature. The entropy production is limited by the third law; the entropy production vanishes as temperature approaches zero. Exception occurs when the ground state exhibits degeneracies either by accidence or due to symmetry. The entropy production then approaches a non-zero finite value completely determined by the number of the degeneracies. Moreover, various non-trivial temperature dependence of the entropy production of the quantum SZE containing bosons or fermions with or without interaction can be understood by considering the third law and such degeneracies. Through this paper we assume that all the measurement is performed in a perfect manner. How the imperfect measurement modifies the results of the classical SZE has been investigated in Ref.~\cite{Sagawa09}. The non-equilibrium effect in the classical SZE has also been considered in Ref. \cite{Sagawa10}.

The rest of this paper is organized as follows. In Sec. II, we introduce the work formula of the quantum SZE and discuss one tricky point of when and how the measurement has been done during the thermodynamic process. Section III is the main part of this paper. We show the role of the third law in the SZE and present the simplest model, one particle SZE. In Sec. IV, we investigate various physical situations of two particle SZE, which shows rich behaviors of a simple information heat engine in the low temperature. Finally, we conclude this paper.

\section{quantum Szilard Engine}

To correctly address the behavior of the SZE in low temperature, we should deal with it in a quantum mechanical way. Recently it has been found by one of us that the work of the quantum SZE is expressed as \cite{Kim11}
\begin{equation}
\left< W \right> = -k_B T \sum_{m=0}^N p_m \ln \left(\frac{p_m}{p^*_m}\right) \equiv k_B T \Delta S,
\label{eq:work}
\end{equation}
where $p_m$ and $p^*_m$ represent the probability to find $m$ atoms in the left side of the box among total $N$ atoms right after a wall is inserted for the time-forward protocol, and that for the time-reversed protocol, respectively. Here  $k_B$, $T$ and $\Delta S$ denote the Boltzmann constant, the temperature of a heat reservoir and the entropy production, respectively.
Recently we reported, however, that Eq.~(\ref{eq:work}) itself can be derived from fully classical consideration. The partition functions, required to calculate $p_n$ and $p^*_n$, differ in whether the classical or the quantum mechanics are taken into account \cite{Kim11a}. All the calculations performed in this paper are based upon Eq.~(\ref{eq:work}).

Now we would like to address one frequently asked question on the SZE: when and how the measurement process has been done during the thermodynamic process. The answer is that the measurement is automatically done by the reservoir when the box is completely separated into two parts by inserting a wall. It is noted that the process of inserting a wall should be performed in an isothermal way in the quantum SZE, which inevitably introduces the reservoir into the problem \cite{Kim11}. In order to make the engine have a well-defined temperature at every moment the reservoir has to observe the eigenenergy of the engine so that each energy eigenstate is properly occupied following the canonical distribution. Below we show that such a measurement of the energy corresponds to that of the information on which part of box the particle is located in when the wall is inserted.

The engine and the wall can be modeled as an one-dimensional box and a potential barrier, respectively; that is,
\begin{eqnarray}
V_{\rm box} = \left\{ \begin{array}{ll}
0 & \textrm{if $0<x<L$}\\
\infty & \textrm{otherwise}
\end{array} \right.
\end{eqnarray}
and
\begin{equation}
V_{\rm wall}(l)=V_0 \delta(x-L/2),
\end{equation}
where $V_0$ is the hight of the wall. For simplicity assume that temperature is negligibly small so that only the ground state is dominantly occupied. When $V_0$ is large enough, the ground and the first excited state, denoted as $|0 \rangle$ and $|1 \rangle$, respectively, are nearly degenerate so as to be almost equally occupied. Thus the density operator is written as
\begin{equation}
\rho = (|0\rangle  \langle 0| + |1\rangle  \langle 1|)/2.
\label{eq:density matrix 1,2}
\end{equation}
Now we introduce $|L \rangle$ ($|R \rangle$), which describes a localized state in the left (right) side of the box, as follows
\begin{eqnarray}
|L \rangle &=& \left( |0 \rangle + |1 \rangle \right)/\sqrt{2} \\
|R \rangle &=& \left( |0 \rangle - |1 \rangle \right)/\sqrt{2}.
\end{eqnarray}
Explicitly $|L \rangle$ is written as
\begin{eqnarray}
|L \rangle = \left \lbrace \begin{array}{ll}
\sqrt{4/L} \sin (2 \pi x /L) & \textrm{if $0<x<L/2$}\\
0 & \textrm{otherwise}
\end{array} \right.
.\end{eqnarray}
$|R \rangle$ can be given in a similar way with the probability localized in the right side. One can then show that Eq.~(\ref{eq:density matrix 1,2}) is mathematically equivalent to
\begin{equation}
\rho=(|L\rangle  \langle L| + |R\rangle  \langle R|)/2.
\label{eq:mixture}
\end{equation}
It means that the state is now represented as the probability mixture of the left and the right side. It is thus no need to perform any additional measurement on which side the particle is in since it has already been done by the reservoir. When the wall is completely inserted the state of the engine is not a coherent superposition of two states $|L \rangle$ and $|R \rangle$ but rather their probability mixture. The similar argument can be found in Ref. \cite{Zurek03}, where once the wall is completely inserted the reservoir is contacted with the engine in order to perform the isothermal expansion. In our case, however, the reservoir should be contacted during the whole process of inserting a wall.

\section{the role of the third law of thermodynamics}

Here we prove that the entropy production, $\Delta S$, is bounded by the entropy of a system, $S$, which guarantees that $\Delta S$ vanishes as $S$ vanishes. When the wall is inserted, the entropy of the system is expressed as
\begin{equation}
S/k_B = -\sum_{m=0}^N \sum_n p_m(n) \ln p_m(n),
\label{eq:entropy of system}
\end{equation}
where $p_m(n)$ represents the probability that the system has a discrete energy $E_n$ when $m$ atoms reside in the left side of the box, and $N-m$ in the right. According to $p_m(n)=p_m p(n|m)$, where $p(n|m)$ is the conditional probability that the system has an energy $E_n$, once $m$ atoms are found in the left side, one finds
\begin{equation}
S/k_B = -\sum_m p_m \ln p_m + \sum_m p_m S_m \ge \Delta S,
\label{eq:inequlaity}
\end{equation}
where $S_m=-\sum_n p(n|m) \ln p(n|m)$. Here $S_m \geq 0$  and $-\ln p_m \ge - \ln (p_m/p^*_m)$ are exploited. As temperature approaches zero, the entropy of system $S$ and accordingly $\Delta S$ approach zero due to the third law.

\begin{figure}
  \includegraphics[width=8.5cm]{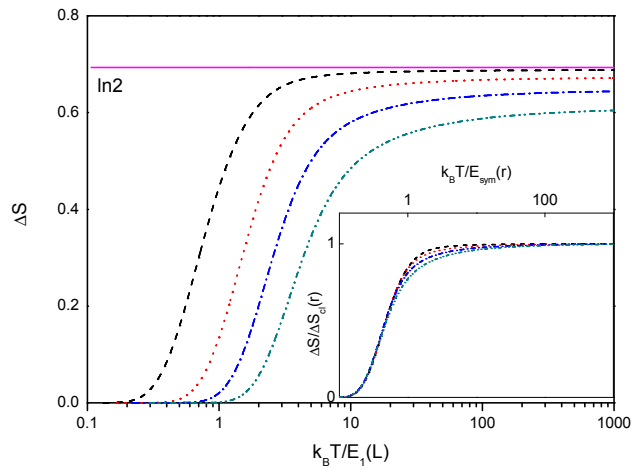}
  \caption{(Color online) $\Delta S$ of the quantum SZE with a single atom as a function of $k_{\rm B}T/E_1(L)$ for $r=0.5$, $0.45$, $0.4$, $0.35$ and $0.3$ from top to bottom, where $E_1(L) = h^2/8mL^2$. (Inset) The same data scaled by $\Delta S_{\rm cl}(r)$ for $y$-aix and $E_{\rm sym}$ for $x$-axis.
  \label{fig2}}
\end{figure}

Now we consider the SZE containing a single atom with the wall inserted at $r=x/L$, where $x$ and $L$ denote the distance between the wall and the left end of the box, and the size of the box, respectively. Here the box is modeled as an infinite potential well. It is shown in Fig.~\ref{fig2} that the entropy productions of the quantum SZE for various $r$ vanish as $T \rightarrow 0$ except $r=0.5$. Note that the entropy production of the classical SZE remains constant, i.e. a binary entropy $\Delta S_{\rm cl}(r) = -r \ln r - (1-r) \ln (1-r)$, independent of temperature. The deviation of the entropy production between the quantum and the classical SZE becomes prominent around $k_B T \sim E_{\rm sym} = \left| E^{L}_1 - E^{R}_1 \right|$, where $E^{L(R)}_n$ denotes the $n$th eigenenergy of the left (right) side of the box ($n=1,2, \cdots$), namely $E^L_n = h^2n^2/8mx^2$ and $E^R_n = h^2n^2/8m(L-x)^2$, where $h$ and $m$ are the Planck constant and a mass of the atom, respectively. The reason is that if $E^{L}_1$ is larger (smaller) than $E^{R}_1$ the left (right) side of the box is predominantly occupied once $k_BT \le E_{\rm sym}$ is satisfied. This is confirmed in the inset of Fig.~\ref{fig2}, where $\Delta S$ scaled by $\Delta S_{\rm cl}(r)$ as a function of $T$ scaled by $E_{\rm sym}$ coalesce with each other for all $r$, and all the curves start to bend at $k_{\rm B}T/E_{\rm sym} \sim 1$.

It should be noted that it was reported in Ref. \cite{Kim11} that the work of the quantum SZE is exactly equivalent to that of the classical one as far as a single atom is concerned. This statement is true only if the wall is inserted precisely at $r=0.5$, at which the ground state exhibits the degeneracy due to the reflection symmetry. The entropy production associated with this degeneracy exactly corresponds to the classical one, namely $\ln 2$. In a generic case, i.e. $r \neq 0.5$, however, the quantum result should deviate from the classical due to the third law.

Let us discuss the meaning of the vanishing entropy production due to the third law in the context of information entropy. As temperature goes to zero, only the ground state becomes dominantly occupied. For $0<r<0.5$, the ground state is located in the right side due to $E^L_1 > E^R_1$. Thus it is no doubt that the atom occupies the ground state of the right side at zero temperature so that the atom should be found in the right without any uncertainty. In the viewpoint of information the entropy is zero since there is no uncertainty for the location of the atom. It also means one cannot transfer any information or entropy to the reservoir by performing measurement and erasure. Therefore, the work cannot be extracted from the engine.

\section{two particle Szilard engine}

If more than one atom is considered, the quantum statistical nature such as bosons and fermions comes into play. However, the third law still governs the low temperature behavior of the entropy production. When the ground state of the system is unique, the entropy production of two atoms vanishes at zero temperature no matter what they are bosons or fermions.

\subsection{Particles without interaction}

\begin{figure}
  \includegraphics[width=8.5cm]{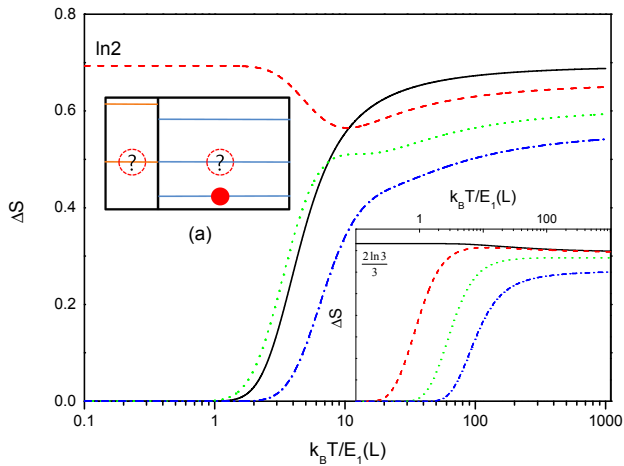}
  \caption{(Color online) $\Delta S$ of the quantum SZE with two non-interacting spinless fermions as a function of $k_{\rm B}T/E_1$ for $r=1/2$ (the solid curve), $1/3$ (the dashed curve), $0.25$ (the dotted curve), and $0.2$ (the dashed dotted curve) and (Inset) two non-interacting spin-$0$ bosons for $r=0.5, 0.45, 0.35$ and $0.25$ from top to bottom. (a) The schematic diagram of energy levels at $r=1/3$. The filled circle represents the ground state is occupied by one fermion while two dotted circles denote two possible ways the second fermion can occupy the next state.
  \label{fig3}}
\end{figure}

In the case of bosons, the degeneracy of the ground state occurs when the reflection symmetry takes place, that is, $E_{\rm sym} = 0$. The vanishing entropy production then becomes prominent for $T \lesssim E_{\rm sym}$, which is clearly shown in the inset of Fig.~\ref{fig3} presenting the entropy production of two non-interacting spin-$0$ bosons. Note here that there exists another energy scale $\Delta E={\rm min} \left\{ E^L_2 - E^L_1, E^R_2 - E^R_1 \right\}$, at which the entropy production starts to increases as $T$ decreases although it is not so dramatic in the inset of Fig.~\ref{fig3}. This rather peculiar behavior is associated with the residual entropy $(2/3) \ln 3$ appearing at $r=0.5$ at $T=0$ \cite{Kim11}, which is larger than the entropy production in the high temperature limit, namely $\ln 2$. As temperature decreases $\Delta S$ first increases around $\Delta E$ and then decreases around $E_{\rm sys}$ (See the dashed curve in the inset of Fig.~\ref{fig3}). Such behavior becomes obscure when $\Delta E \lesssim E_{\rm sym}$ (See the dashed dotted curve in the inset of Fig.~\ref{fig3}).

Two fermions exhibit completely different behavior. It is shown in Fig.~\ref{fig3} that the entropy production of two non-interacting {\em spinless} fermions approaches zero even for $r=1/2$. The reason is that the ground state of two fermions is non-degenerate due to the Pauli's exclusion principle \cite{Kim11}. However, at $r=1/3$ as temperature decreases the entropy production first decreases but finally approaches $\ln 2$ (See the dashed curve in Fig.~\ref{fig3}). This unexpected behavior is ascribed to the accidental degeneracy between the ground state of the left side and the first excited state of the right as shown in Fig.~\ref{fig3}(a). One fermion then occupies the ground state of the right while the second fermion is allowed to occupy either the left or the right, which contributes to entropy by $\ln 2$.

\subsection{Particles with interaction}

Let us now consider more realistic situations, namely two spin-zero bosons with either attractive or repulsive interaction and two spin-$1/2$ fermions with either ferromagnetic or anti-ferromagnetic interaction. It is not surprising that the results will become more complicated. However, the basic principle is the same as before: how the entropy production behaves in the low temperature limit is completely determined by the third law. Now we have one more energy scale, i.e. the interaction energy. Make sure that the work can still be expressed as Eq.~(\ref{eq:work}) even if the interaction between atoms takes place.

\begin{figure}
  \includegraphics[width=8.5cm]{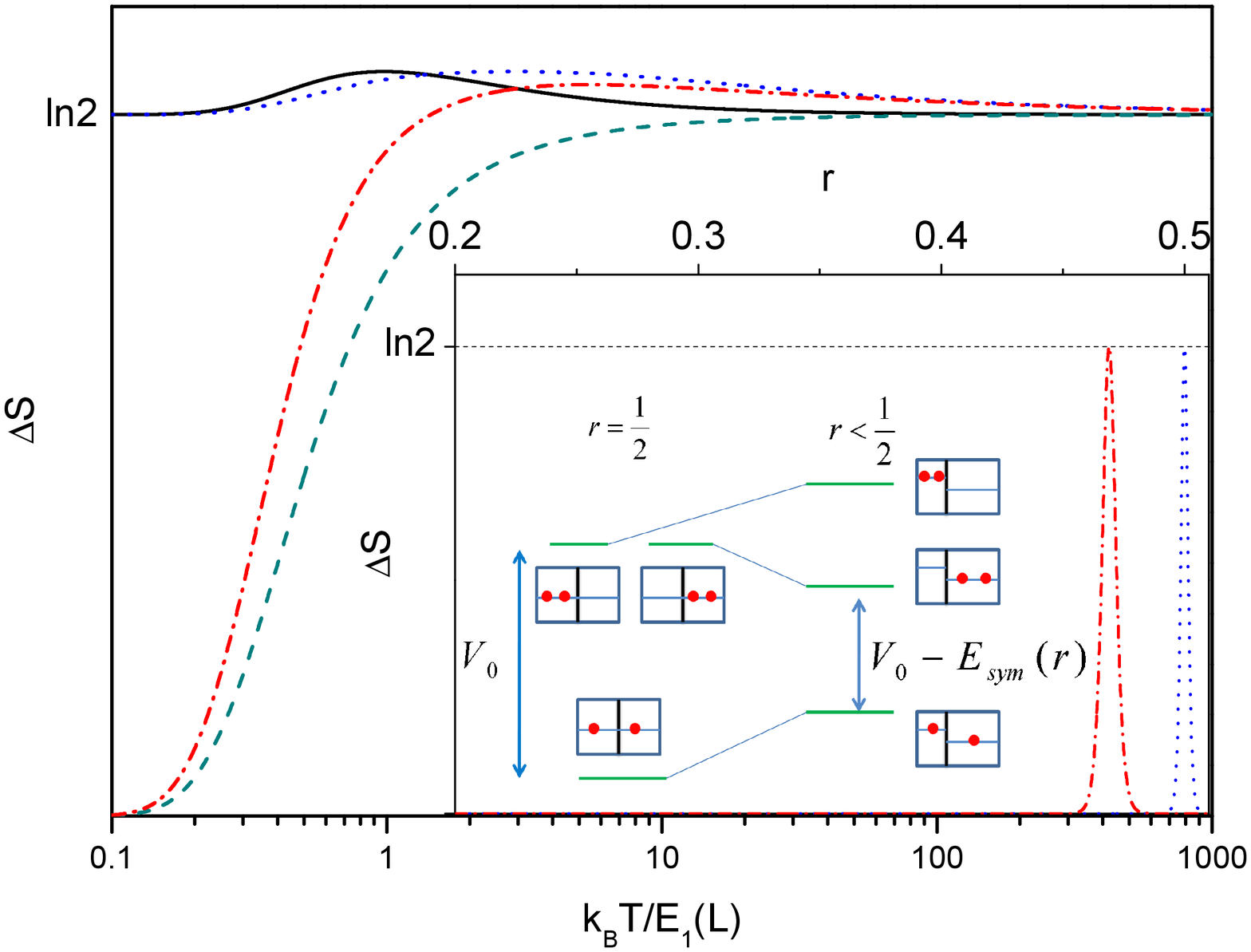}
  \caption{(Color online) $\Delta S$ of the classical SZE of two distinguishable particles with the attractive (the solid curve) and the repulsive interaction (the dashed curve) and of the quantum SZE of two spin-$0$ bosons with the attractive (the dotted curve) and the repulsive interaction (the dashed-dotted curve) as a function of $k_{\rm B}T/E_1$ for $r=1/2$. (Inset) $\Delta S$ as a function of $r$ at $k_{\rm B}T/E_1=0.05$. The schematic diagram represents the energy levels of several configurations depending on $r$ (See the text for the detail).
  \label{fig4}}
\end{figure}

\subsubsection{Spin-0 bosons with attractive or repulsive inertaction}

Here two interacting bosons are taken into account. For simplicity the interaction potential $V$ for two bosons is modeled as follows: $V=V_0$, where $V_0$ is negative (positive) for the attractive (repulsive) interaction, if two bosons are at the same side, otherwise $V=0$. For the attractive interaction there exist two degenerate ground states for $r=1/2$ at zero temperature giving rise to $\Delta S = \ln 2$ since two atoms behave like a molecular pair by residing in the same side. This is indeed equivalent to the single atom case studied above, and becomes pronounced when $k_B T \lesssim |V_0|$. The slight increase of $\Delta S$ larger than $\ln 2$ (The dotted curve of Fig.~\ref{fig4}) is a classical effect (See the solid curve of Fig.~\ref{fig4}). Even in the classical case it is more likely for two atoms to reside at the same side when the attractive interaction exists, from which configuration one can extract more entropy production than a non-interacting case. For the repulsive interaction, the residual entropy at zero temperature is zero since for $k_{\rm B}T \lesssim V_0$ two bosons considerably repel each other like two spinless fermions discussed above (See the dashed-dotted curve of Fig.~\ref{fig4}). The similar behavior can also be observed in classical case (The dashed curve of Fig.~\ref{fig4}). Note that these correspondences between the quantum and the classical SZE are available only at $r=1/2$. As $r$ decreases from $1/2$, the entropy production at $k_{\rm B}T/E_1=0.05$ rapidly vanishes for the attractive potential as shown in the inset of Fig.~\ref{fig4}.
For the repulsive potential, however, the maximum occurs at $r \sim 0.47$ since the accidental degeneracy between two configurations shown schematically in the inset of Fig.~\ref{fig4} can occur at $V_0 = E_{\rm sym}$.

\subsubsection{Spin-1/2 fermions with anti-ferromagnetic or ferromagnetic interaction}

\begin{figure}
  \includegraphics[width=8.5cm]{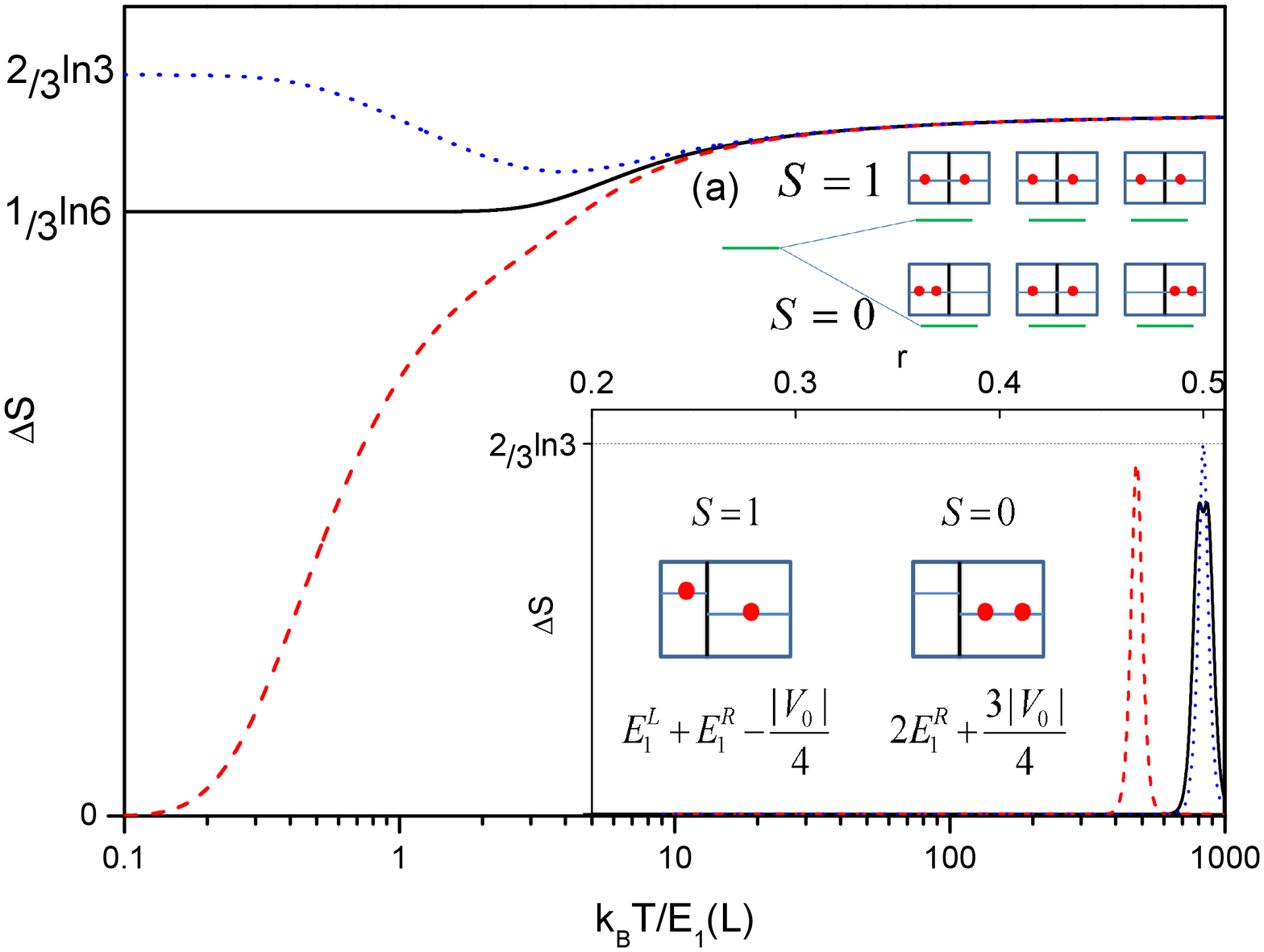}
  \caption{(Color online) $\Delta S$ of the quantum SZE with two spin-$1/2$ fermions with no interaction (the solid curve), the ferromagnetic (the dashed curve) and the anti-ferromagnetic (the dotted curve) interaction as a function of $k_{\rm B}T/E_1$ for $r=1/2$. (a) The schematic diagram represents the energy levels of the spin singlets ($S=0$) and the triplets ($S=1$). (Inset) $\Delta S$ as a function of $r$ at $k_{\rm B}T/E_1=0.05$.
  \label{fig5}}
\end{figure}

Now let us consider two spin-$1/2$ fermions with spin-spin interaction modeled as
\begin{equation}
V=V_0 \vec{s}_1 \cdot \vec{s}_2 /\hbar^2,
\end{equation}
where $\vec{s}_1$ and $\vec{s}_2$ represent the spin of each fermion, and $V_0$ is a positive (negative) constant for anti-ferromagnetic (ferromagnetic) interaction. The interaction potential can be rewritten as
\begin{equation}
V=V_0 (S^2/2\hbar^2 -3/4)
\end{equation}
with $\vec{S}=\vec{s}_1+\vec{s}_2$. If no interaction is considered, there exist six degenerate ground states for $r=1/2$ at zero temperature; namely $\psi_L(1)\psi_L(2) \left| 0,0 \right>$, $\psi_R(1)\psi_R(2) \left| 0,0 \right>$, $ \psi_+\left| 0,0 \right>$, $\psi_-\left| 1,1 \right>$, $\psi_-\left| 1,0 \right>$ and $\psi_-\left| 1,-1 \right>$, where $\psi_{L(R)}$ and $\left| S,M \right>$ denote the wavefunctions of the coordinate space in the left (right) side and those of the total spin states, respectively, and $\psi_{\pm}$ represents $(1/\sqrt{2}) [\psi_L(1)\psi_R(2) \pm \psi_L(2)\psi_R(1)]$. It is noted that the work can be extracted only from two of them, that is, $\psi_L(1)\psi_L(2) \left| 0,0 \right>$ and $\psi_R(1)\psi_R(2) \left| 0,0 \right>$, since the others have one fermion per each side giving rise to no pressure difference between two sides. This explains the entropy production approaches $(1/3)\ln 6$ rather than $\ln 6$ at zero temperature as shown in Fig.~\ref{fig5}.

As far as the anti-ferromagnetic interaction is concerned, six degenerate levels are split into two groups so that the energy of the spin singlets ($S=0$), namely $-3V_0/4$, becomes smaller than that of the triplets ($S=1$), namely $V_0/4$, as shown in Fig.~\ref{fig5}(a). For $k_{\rm B}T \lesssim V_0$, there exist three degenerate ground states instead of six. It gives rise to $\Delta S = (2/3) \ln 3$, where $2/3$ reflects that the work is extractable from two cases among three [See the lower row of Fig.~\ref{fig5}(a)]. For the ferromagnetic interaction, on the other hand, the opposite occurs; that is, the energy of the singlets become larger than that of the triplets. Although again the degeneracy of the ground states at zero temperature is still three, the extractable work vanishes since there exists one fermion per one side in every configuration [See the upper row of Fig.~\ref{fig5}(a)]. When the reflection symmetry is broken, i.e. $r\neq 1/2$, the entropy production $k_{\rm B}T/E_1=0.05$ rapidly vanishes for both the non-interacting and the anti-ferromagnetic cases as shown in the inset of Fig.~\ref{fig5}.

For the ferromagnetic case, however, the maximum occurs at $r \sim 0.47$. This might be associated with the accidental degeneracy similar to that shown in the inset of Fig.~\ref{fig4}. However, the story is not that simple. The accidental degeneracy between two configurations, namely $S=0$ and $S=1$, presented schematically in the inset of Fig.~\ref{fig5}, indeed occurs at $r \sim 0.47$. Since $S=1$ is triple-degenerate while $S=0$ non-degenerate, the entropy production of such configuration is given as
\begin{equation}
\Delta S = -(1/4) \ln (1/4)- (3/4) \ln (3/4).
\label{eq:entropy_r_deg}
\end{equation}
However, this is smaller than $\ln2$, the maximum value of $\Delta S(r)$.

\begin{figure*}[ht]
\begin{center}
  \includegraphics[width=15cm]{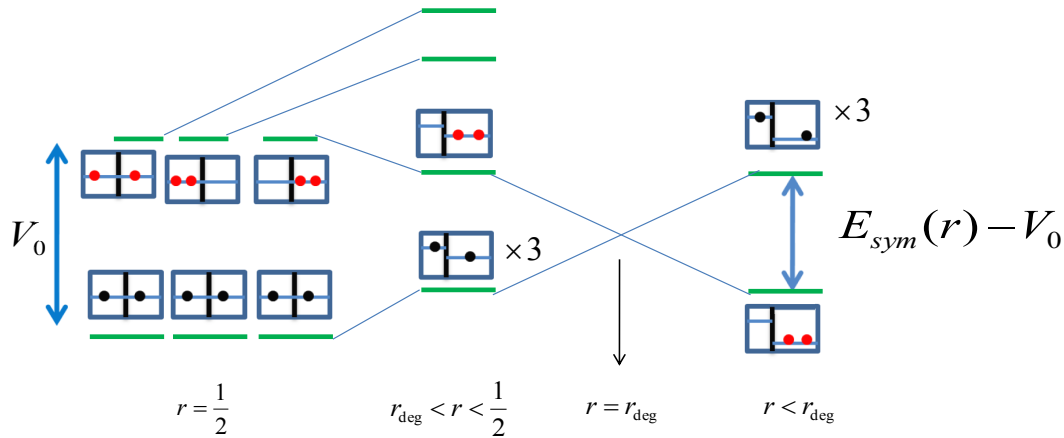}
\end{center}
  \caption{Schematic diagram of the energy levels depending on r. $V_0/E_1(L) = -1$ is taken leading to $r_{\rm deg} \sim 0.47$.  \label{figS1}}
\end{figure*}

Figure~\ref{figS1} schematically shows the configurations of energy levels with $r$ varied. The degeneracy between $S=0$ and $S=1$ takes place at $r=r_{\rm deg}$ which satisfies $\delta E(r_{\rm deg}) \equiv E_{\rm sym} (r_{\rm deg}) - V_0 =0$. By using Eq.~(1) the entropy production can be written as
\begin{equation}
\Delta S = -p_0 \ln p_0- p_1 \ln p_1,
\label{eq:entropy production}
\end{equation}
where $p^*_0= 1$ and $p^*_1 \approx 1$ in zero temperature limit are exploited. Here we can safely assume $p_2 \approx 0$, implying $p_0+p_1 \approx 1$ due to $p_2/p_m \sim e^{-\Delta \epsilon_m/k_B T} \ll 1$ as $T \rightarrow 0$, in which $\Delta \epsilon_m > 0$ denotes the energy difference between the configurations of two fermions and $m$ fermions in the left side ($m=0, 1$). Thus $p_0$ and $p_1$ are given as
\begin{equation}
p_0=\frac{1}{3 e ^{-\delta E /k_BT}+1}
\end{equation}
and
\begin{equation}
p_1=\frac{3 e ^{-\delta E /k_BT}}{3 e ^{-\delta E /k_BT}+1},
\end{equation}
where $3$ represents triple-degeneracy of $S=1$ state. Once we have $\delta E (r_{\rm deg})=0$, Eq.~(\ref{eq:entropy_r_deg}) is recovered.

\begin{figure}
  \includegraphics[width=8.5cm]{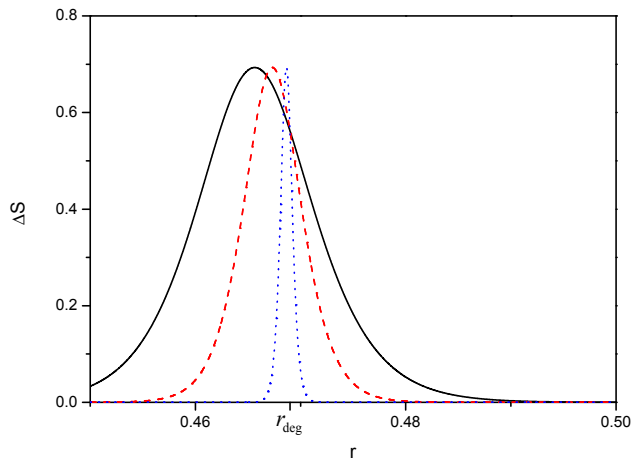}
  \caption{$\Delta S$ as a function of r for $k_BT/E_1(L)=0.1$ (the solid curve), $0.05$ (the dashed curve) and $0.01$ (the dotted curve). \label{figS2}}
\end{figure}

As a matter of fact, Eq.~(\ref{eq:entropy production}) is maximized by $\ln 2$ when $p_0=p_1=1/2$ or equivalently $\delta E /k_BT = \ln 3$ is satisfied. It looks strange that the entropy production of the non-degenerate ground state is not equal to zero, and moreover exhibits the maximum possible value. This is ascribed to finite temperature effect. For a non-zero temperature, no matter how small it is, there exist $r^*$ maximizing $\Delta S$ by $\ln 2$ once $\delta E /k_BT = \ln 3$ is fulfilled. As $T \rightarrow 0$, $r^*$ approaches $r_{\rm deg}$ as shown in Fig.~\ref{figS2}. Therefore, the maximum value of $\Delta S$, $\ln 2$, in the inset of Fig.~\ref{fig5} is correct, and it occurs not at $r=r_{\rm deg}$ but at $r=r^*$.

\section{Conclusion}
In summary, we have shown that the third law of thermodynamics plays an important role in understanding the low temperature behavior of the SZE. When there is no degeneracy in the ground state the entropy production vanishes in the zero temperature limit. Various degeneracies generated from quantum statistical nature, symmetries, interaction, and accidence determine the entropy production in the low temperature, which limits the work extractable from the SZE.

We would like to thank Takahiro Sagawa, Simone De Liberato, and Masahito Ueda for useful discussions. This was supported by the NRF grant funded by the Korea government (MEST) (No.2009-0084606, No.2009-0087261 and No.2010-0024644).

\end{document}